\documentclass[pre,showpacs,amssymb]{revtex4}
\usepackage{amsmath,graphicx}
\usepackage{epsfig}

\begin{document}

\title{Pulling an adsorbed polymer chain off a solid surface}

\author{S. Bhattacharya$^1$, A. Milchev$^{1,2}$, V.G. Rostiashvili$^1$ and T.A.
Vilgis$^1$} \affiliation{$^1$ Max Planck Institute for Polymer Research 10
Ackermannweg, 55128 Mainz, Germany\\
$^2$ Institute for Physical Chemistry, Bulgarian Academy of Science, 1113 Sofia,
Bulgaria}

\begin{abstract}
The thermally assisted detachment of a self-avoiding polymer chain from an
adhesive surface by an external force applied to one of the chain ends is
investigated. We perform our study in the ``fixed height'' statistical ensemble
where one measures the fluctuating force, exerted by the chain on the last
monomer when a chain end is kept fixed at height $h$ over the solid plane at
different adsorption strength $\epsilon$. The phase diagram in the $h -
\epsilon$ plane is derived both analytically and by Monte Carlo simulations.
We demonstrate that in the vicinity of the polymer desorption transition a
number of properties like fluctuations and probability distribution of various
quantities  behave differently, if $h$ rather than $f$ is used as an independent
control parameter.
\end{abstract}
\maketitle

\section{Introduction}\label{Intro}
The properties of single polymer chains at surfaces have received considerable
attention in recent years. Much of this has been spurred by new experimental
techniques such as atomic force microscopy (AFM) and optical/magnetic tweezers
\cite{Smith} which allow one to manipulate single polymer chains. Study of
single polymer molecules at surfaces, such as mica or self-assembled monolayers,
by Atomic Force Microscopy (AFM) method provides a great scope for
experimentation \cite{Senden,Hugel,Seitz,Rohrbach,Gaub_2,Friedsam,Gaub_1}.
Applications range from sequential unfolding of collapsed biopolymers over
stretching of coiled synthetic polymers to breaking individual covalent
bonds\cite{Rief,Ortiz,Grandbois}.

In these experiments it is customary to anchor a polymer molecule with one end
to the substrate whereas the other end is fixed on the AFM cantilever. The
polymer molecule can be adsorbed on the substrate while the cantilever recedes
from the substrate. In so doing one can prescribe the acting force in AFM
experiment whereas the  distance between the tip and the surface is measured.
Conversely, it is also possible to fix the distance and measure the
corresponding force, a method which is actually more typical in AFM-experiments.
From the standpoint of statistical mechanics these two cases could be qualified
as $f$-ensemble (force is fixed while fluctuating the height of the chain end is
measured) and as $h$-ensemble ($h$ is fixed while one measures the fluctuating
$f$). Recently these two ways of descriptions as well as their interrelation
were discussed for the case of a phantom polymer chain by Skvortsov et al.
\cite{Skvortsov}. In recent papers \cite{PRER,Bhattacharya} we studied
extensively the desorption of a single tethered {\em self-avoiding} polymer from
a flat solid substrate with an external force applied to a free chain's end in
the $f$-ensemble.

In the present paper we consider the detachment process of a single
self-avoiding polymer chain, keeping the distance $h$ between the free chain's
end and the substrate as the control parameter. We derive analytical results for
the main observables which characterize the detachment process. The mean
value as well as the probability distribution function (PDF) of the order
parameter are presented in close analytical expressions using the
Grand Canonical Ensemble (GCE) method\cite{Bhattacharya}. The basic force-height
relationship which describes the process of polymer detachments by pulling, that
is, the relevant equation of state for this system is also derived both for
extendible as well as for rigid bonds and shown to comply well with our Monte
Carlo simulation results. We demonstrate also that a number of properties behave
differently in the vicinity of the phase transition, regarding which of the two
equivalent ensembles is used as a basis for the study of systems's behavior.

\section{Single chain adsorption: using distance as a control parameter}
\label{Theory}
\subsection{Deformation of a tethered chain}

Before considering the adsorption-desorption behavior of a polymer in terms of
the chain end distance $h$, we first examine how a chain tethered to a solid
surface responds to stretching. This problem amounts to finding the chain free
end probability distribution function (PDF) $P_{N}(h)$ where $N$ is the chain
length, i.e., the number of beads. The partition function of such a chain at
fixed distance $h$ of the chain end from the anchoring plane is given by
\begin{eqnarray}
 \Xi_{\rm tail}(N, h) = \frac{\mu_{3}^{N}}{N^{\beta}} \: l_0 P_{N} (h)
\label{Partition_Tail}
\end{eqnarray}
where $\beta = 1 - \gamma_1$ and the exponent $\gamma_1 =
0.680$\cite{Vanderzande}. Here $\mu_3$ is a model dependent connective constant
(see e.g. ref.\cite{Vanderzande}). In Eq.~(\ref{Partition_Tail}) $l_0$ denotes a
short-range characteristic length which depends on the chain model. Below we
discuss the chain deformation within two models: the bead-spring (BS) model for
elastic bonds and the freely jointed bond vectors (FJBV) model where the bonds
between adjacent beads are considered rigid.

\subsubsection{Bead-spring model}

The form of $P_N(h)$ has been discussed earlier \cite{Binder,Eisenriegler} and
later used in studies of the monomer density in polymer brushes \cite{Kreer}.
Here we outline this in a way which is appropriate for our purposes.  The
average end-to-end chain distance scales as $R_N = l_0 N^{\nu}$, where $l_0$ is
the mean distance between two successive beads on a chain and $\nu \approx
0.588$ is the Flory exponent \cite{Grosberg}.  The short distance behavior, $h
\ll R_N$, is given by
\begin{eqnarray}
 P_N(h) \propto \left(\frac{h}{R_N}\right)^{\zeta}
\label{Short}
\end{eqnarray}
where the exponent $\zeta \approx  0.8$. For the long distance
behavior, $h/R_N \gg 1$,  we assume,
following Ref.\cite{Kreer}, that the PDF of the end-to-end vector ${\bf r}$ is
given by des Cloizeaux's expression \cite{Cloiseaux} for a chain in the bulk :
$P_N({\bf r})=(1/R_N) \: F({\bf r}/R_N)$ where the scaling function $F(x)
\propto x^t \: \exp [-D x^\delta]$, and the exponents $t=(\beta-d/2+\nu
d)/(1-\nu)$ , $\delta=1/(1-\nu)$. Here and below $d$ denotes the space
dimensionality. One should emphasize that the presence of a surface is
manifested only by the replacement of the universal exponent $\gamma$ with  another
universal exponent $\gamma_1$ (as compared to the pure bulk case!). By
integration of $P_N({\bf r})$ over the $x$ and $y$ coordinates while $h$ is
measured along the $z-$coordinate, one obtains $P_N(h) \propto
(h/R_N)^{2+t-\delta} \:\: \exp [-D(h/R_N)^{\delta}]$. As the long distance
behavior is dominated mainly by the exponential function while the short
distance regime is described by Eq.~(\ref{Short}), we can approximate the
overall behavior as
\begin{eqnarray}
 P_N(h)  = \frac{A}{R_N} \: \left(\frac{h}{R_N}\right)^{\zeta} \: \exp \left[-D
\left(\frac{h}{R_N}\right)^{\delta}\right]
\label{Overall}
\end{eqnarray}
\vspace*{0.9cm}
\begin{figure}[htb]
\includegraphics[scale=0.34]{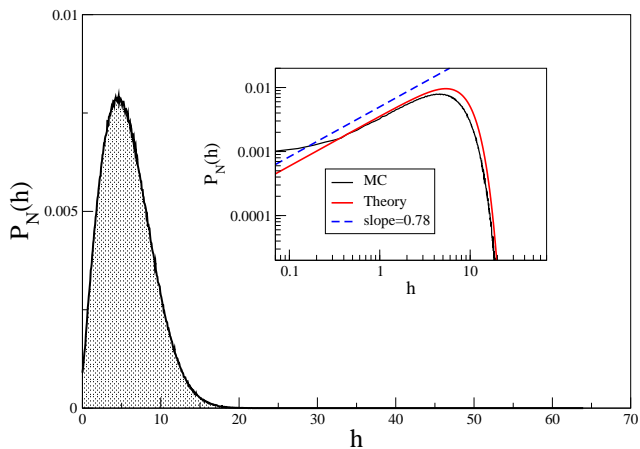}
\caption{Probability distribution $P_N(h)$ of chain end positions $h$ above the
grafting plane for a polymer with $N = 128$ monomers at zero strength of the
adsorption potential $\epsilon = 0.0$. In the inset the MC data for $P_N(h)$
(solid black line) is compared to the theoretic result, Eq.~(\ref{Overall}).
Dashed line denotes the expected slope of $\zeta \approx 0.78$ of the
probability distribution for small heights.}
\label{pdf_ends}
\end{figure}
A comparison of the distribution, Eq.~(\ref{Overall}), with our simulation data
is shown in Fig.~\ref{pdf_ends}. The constants $A$ and $D$ in
Eq.~(\ref{Overall}) can be found from the conditions: $\int P_N (h) d h =1$ and
$\int h^2 P_N (h) d h =R_N^2$. This leads to:
\begin{eqnarray}
 A = \delta
\left[\Gamma\left(\frac{1+\zeta}{\delta}\right)\right]^{-(1+\zeta)/2}
\left[\Gamma\left(\frac{3+\zeta}{\delta}\right)\right]^{-(1-\zeta)/2}
\label{A}
\end{eqnarray}
and
\begin{eqnarray}
 D = \left[\Gamma\left(\frac{3+\zeta}{\delta}\right)\right]^{\delta/2}
\left[\Gamma\left(\frac{1+\zeta}{\delta}\right)\right]^{-\delta/2}
\label{D}
\end{eqnarray}
where $\delta \approx 2.43$ and $\zeta \approx 0.8$. One gets thus the estimates
$A\approx 2.029$ and $D \approx 0.670$.

The free energy of the tethered chain with a fixed distance $h$ takes on the
form $F_{\rm tail}(N, h) = - k_BT \ln \Xi_{\rm tail} (N, h)$ where $k_B$
denotes the Boltzmann constant. By making use of Eqs. (\ref{Partition_Tail}) and
(\ref{Overall}) the expression for the force $f_N$, acting on the end-monomer
when kept at distance $h$ is given by
\begin{eqnarray}
 f_N = \frac{\partial}{\partial h} \: F_{\rm tail}(N, h)
= \frac{k_BT}{R_N} \:\left[ \delta D \left( \frac{h}{R_N}\right)^{\delta-1} -
\zeta \left( \frac{R_N}{h}\right)\right]
\label{Deformation}
\end{eqnarray}
One should note that at $h/R_N \gg 1$ we have $h \propto R_N (R_N
f_N/k_BT)^{1/(\delta-1)}$ which, after taking into account that $\delta^{-1} =
1-\nu$, leads to the well known Pincus deformation law: $h \propto l_0 N (l_0
f_N/k_BT)^{1/\nu -1}$ \cite{de Gennes}. Within the framework of this
approximation the (dimensionless) elastic energy reads $U_{\rm el}/k_BT = - N
(l_0 f_{N}/k_BT)^{1/\nu}$. In result the corresponding free energy of the chain
tail is given by
\begin{eqnarray}
 \frac{F_{\rm tail}}{k_BT} =   - N \left(\frac{l_0 f_N}{k_BT}\right)^{1/\nu} - N
\ln \mu_3
\label{Free_Energy_BS}
\end{eqnarray}

Eq.~(\ref{Deformation}) indicates that there
exists a height $h_0 = (\zeta/\delta D)^{1/\delta} R_N$ over the surface where
the force $f_N$ changes sign and becomes negative (that is, an entropic
repulsion dominates). According to Eq.~(\ref{Deformation}) the force diverges as
$f_N \propto - k_BT/h$ upon further decrease of the distance $h$.

\subsubsection{Freely jointed chain}

It is well known \cite{Grosberg} that the Pincus law, Eq. (\ref{Deformation}),
describes the deformation of a linear chain at intermediate force strength,
$1/N^{\nu} \ll l_0 f_N/k_BT \leq 1$. Direct Monte Carlo simulation results
indicate that, depending on the model, deviations from Pincus law emerge at
$h/R_N \ge 3$ (bead-spring off-lattice model) \cite{Lai}, or $h/R_N \ge 6$ (Bond
Fluctuation Model \cite{Wittkop}). In such ``overstretched'' regime (when the
chain is stretched close to its contour length) one should take into account
that the chain bonds cannot expand indefinitely. This case could be treated,
therefore, within the simple freely jointed bond vectors (FJBV) model
\cite{Lai,Schurr} where the bond length $l_0$ is fixed. In this model the force
- deformation relationship is given by
\begin{eqnarray}
 f_N = \frac{k_BT}{l_0} \: {\cal L}^{-1} \left( \frac{h}{l_0 N}\right)
\label{L}
\end{eqnarray}
where ${\cal L}^{-1}$ denotes the inverse Langevin function ${\cal L}(x) =
\coth(x) - 1/x$ and $l_0$ is the fixed bond vector length.  We discuss the main
results pertaining to the FJBV model in Appendix A. The elastic deformation
energy reads $U_{\rm el}/k_BT = - (l_0 f_N/k_BT)  \sum_{i=1}^{N} <\cos \theta_i>
= - N (l_0 f_N/k_BT) {\cal L}(l_0 f_N/k_BT)$, where $\theta_i$ is the average
polar angle of the $i$-th bond vector (see Appendix A). Thus the corresponding
free energy of the chain tail for the FJBV model reads
\begin{eqnarray}
 \frac{F_{\rm tail}}{k_BT} =   - N {\cal G}\left(\frac{l_0 f_N}{k_BT} \right)  -
N \ln \mu_3
\label{Free_Energy_FJBV}
\end{eqnarray}
where we have used the notation ${\cal G}(x) = x {\cal L}(x) = x \coth (x) - 1$.
Now we are in a position to discuss the pulling of the adsorbed chain controlled
by the chain height $h$.

\subsection{Pulling controlled by the chain end position}

\begin{figure}[bht]
\hspace{-2.0cm}
\includegraphics[scale=0.6]{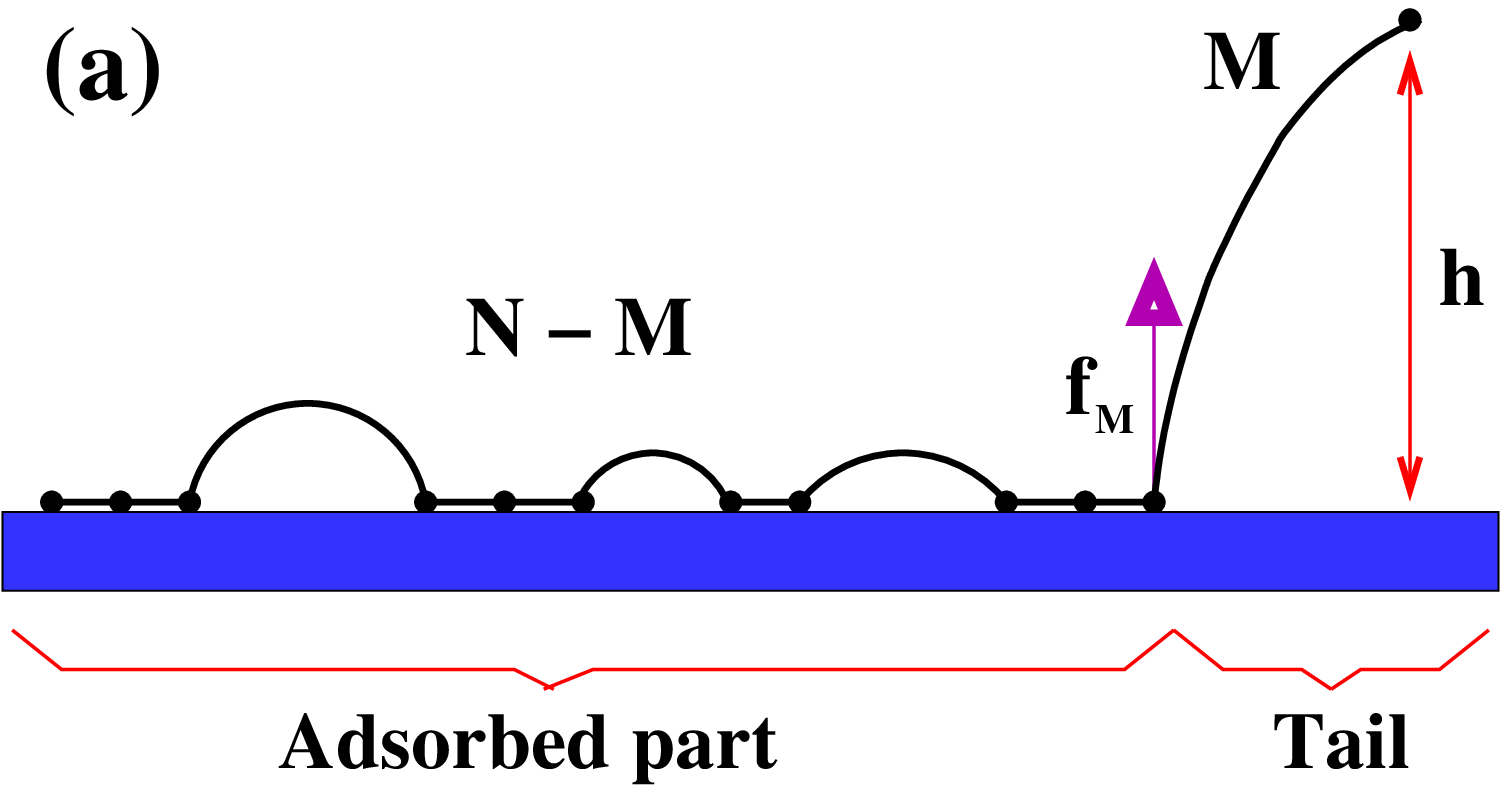}
\hspace{0.5cm}
\includegraphics[scale=0.3]{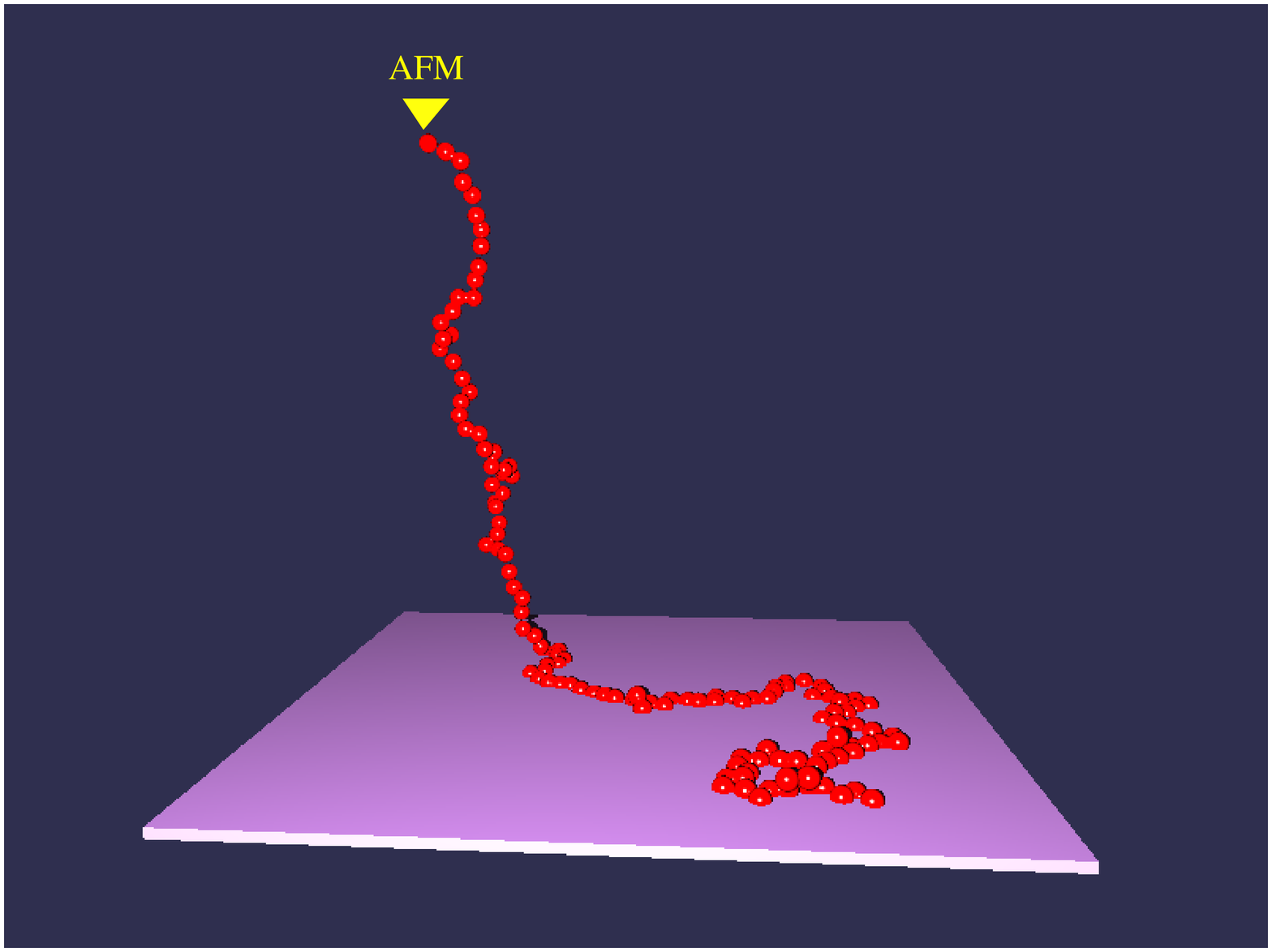}
\caption{ (a) Schematic graph of an adsorbed polymer chain, partially detached
from the plane by an external force which keeps the last monomer at height $h$.
The total chain  is built up from a tail of length $M$ and an adsorbed part
of length $N - M$. The force $f_M$ acting on the chain end is conjugated to $h$,
i.e., $f_M = \partial F_{\rm tail}/\partial h$. (b) A snapshot from the MC
simulation: $N = 128,\; h = 25.0,\; \epsilon = 4.0$ and $\langle f \rangle =
6.126$. }
\label{Tearing}
\end{figure}

Consider now an adsorbed chain when the adsorption energy per monomer is
sufficiently large, $\varepsilon \geq \varepsilon_c$, where $\varepsilon_c$
denotes a corresponding {\it critical} energy of adsorption. Below we will also
use the notation $\epsilon = \varepsilon/k_B T$ for the dimensionless adsorption
energy. The problem of force-induced polymer desorption could be posed as
follows: how is the process of polymer detachment governed by the chain end
position $h$? Figure \ref{Tearing}a gives a schematic representation of such a
system, and the situation in a computer experiment, as shown in the snapshot
Fig.~\ref{Tearing}b, is very similar.

As is evident from Fig. \ref{Tearing}a, the system is built up from a tail of
length $M$ and an adsorbed portion of length $N-M$. The adsorbed part can be
treated within  the GCE approach \cite{Bhattacharya}. In our earlier treatment
\cite{Bhattacharya} it was shown that the free energy of the adsorbed
portion is $F_{\rm ads} = k_BT (N - M)\ln z^{*}(\epsilon)$, where the fugacity
per adsorbed monomer $z^{*}(\epsilon)$ depends on $\epsilon$ and can be found
from the basic equation
\begin{eqnarray}
 \Phi(\alpha, \mu_3 z^{*}) \: \Phi (\lambda, \mu_2 w z^{*})= 1
\label{Basic}
\end{eqnarray}
The so called {\it polylog function} in Eq. (\ref{Basic}) is defined as $\Phi
(\alpha, z) = \sum_{n=1}^{\infty} z^n/n^{\alpha}$ and the connective constants
$\mu_3$, $\mu_2$ in three and two dimensional space have values which are model
dependent \cite{Vanderzande}.  The exponents $\alpha = 1+ \phi$ and $\lambda =
1-\gamma_{d=2}$ where $\phi \approx 0.5$ is the {\it crossover exponent} which
governs the polymer adsorption at criticality, and in particular, the fraction
of adsorbed monomers at the critical adsorption point (CAP) $\epsilon =
\epsilon_{c}$. The constant $\gamma_{d=2} = 1.343$ \cite{Vanderzande}. Finally
$w = \exp(\epsilon)$ is the additional statistical weight gained by each
adsorbed segment.

In equilibrium, the force conjugated to $h$, that is, $f_M = \partial F_{\rm
tail}/\partial h$, should be equal to the {\it chain resistance force to
pulling} $f_{\rm p} = (k_BT/l_0) {\cal F}(\epsilon)$ (where ${\cal F}(\epsilon)$
is a scaling function depending only on $\epsilon$), i.e.,
\begin{eqnarray}
 f_M = \begin{cases}\frac{k_BT}{R_M} \:\left[ \delta D \left(
\frac{h}{R_M}\right)^{\delta-1} -
\zeta \left( \frac{R_M}{h}\right)\right] = f_{\rm p}  &\mbox{, for BS-model}\\
\\
\frac{k_BT}{l_0} \: {\cal L}^{-1}\left(\frac{h}{l_0 M}\right) =  f_{\rm p}
&\mbox{,
for FJBV-model}
\end{cases}
\label{Mech_Euqilibrium}
\end{eqnarray}
The resisting force $f_{\rm p}$ holds the last adsorbed monomer on the adhesive
plane (see again Fig.\ref{Tearing}a  whereby this monomer is shown to experience
a force $f_{M}$). One should emphasize that the force $f_{\rm p}$ {\em stays
constant} in the course of the pulling process as long as one monomer, at least,
is adsorbed on the surface. Thus  $f_{\rm p}$ corresponds to a {\it plateau} on
the deformation curve (force $f$ vs. chain end position $h$). The adsorbed
monomer (see Fig. \ref{Tearing}) has a chemical potential, $\mu_{\rm ads}=\ln
z^{*}$, which in equilibrium should be equal to the chemical potential of a
desorbed monomer in the tail, $\mu_{\rm des}= \partial (F_{\rm
tail}/k_BT)/\partial N$. The expression for $F_{\rm tail}$ depends on the model
and is given either by Eq.(\ref{Free_Energy_BS}) for the BS-model or by
Eq.(\ref{Free_Energy_FJBV}) in the case of FJBV-model. Taking this into account
the condition $\mu_{\rm ads}=\mu_{\rm des}$ leads to the following ``plateau
law'' relationship
\begin{eqnarray}
 \frac{l_0 \: f_{\rm p}}{k_BT} = \begin{cases}\left|\ln [\mu_3
z^{*}(\epsilon)]\right|^{\nu} &\mbox{, for BS-model}\\
\\
{\cal G}^{-1}\left(\left|\ln [\mu_3 z^{*} (\varepsilon)]\right|\right) &\mbox{,
for
FJVB-model}
\end{cases}
\label{Local_Equilibrium}
\end{eqnarray}
where ${\cal G}^{-1}(x)$ stands for the inversion of the function ${\cal G}(x) =
x \coth (x) - 1$.  One should note that Eq.(\ref{Local_Equilibrium}) coincides
with Eq.(3.16) in Ref. \cite{Bhattacharya} which determines the detachment line
in the pulling process controlled by the applied force. Close to the critical
point $\epsilon_c$, the plateau force $f_{\rm p}$ goes to zero. Indeed, since in
the vicinity of the critical point $\ln [\mu_3 z^{*}(\epsilon)] \propto -
(\epsilon - \epsilon_c)^{1/\phi}$ (see ref.\cite{Bhattacharya}), and ${\cal
G}^{-1} (x)\approx (3 x)^{1/2}$, one may conclude that $f_{\rm p} \propto
(\epsilon - \epsilon_c)^{\nu/\phi}$ for the BS-model and $f_{\rm p} \propto
(\epsilon - \epsilon_c)^{1/2 \phi}$ for the FJVB-model.

One can solve Eq.(\ref{Mech_Euqilibrium}) with respect to $M$ (taking into
account that $h \gg R_M$), and arrive at an expression for the tail length
\begin{eqnarray}
 M(h, \epsilon) = \begin{cases} \frac{h}{l_0} \: \left(\frac{k_BT}{l_0
f_{\rm p}}\right)^{1/\nu - 1} &\mbox{, for BS-model}\\
\\
\frac{h}{l_0} \:\left[{\cal L}\left(\frac{l_0
f_{\rm p}}{k_BT}\right)\right]^{-1}&\mbox{, for
FJVB-model}
\end{cases}
\label{M_vs_h}
\end{eqnarray}
where the force at the plateau, $f_{\rm p}$, is described by Eq.
(\ref{Local_Equilibrium}). If for the degree of adsorption one uses as an order
parameter the fraction of chain contacts with the plane, $n = N_s/N$, where
$N_s$ is the number of monomers on the surface, one can write
\cite{Bhattacharya}
\begin{eqnarray}
n = - \frac{1}{k_BT N} \frac{\partial}{\partial \epsilon} \left(F_{\rm ads} +
F_{\rm tail}\right)
\label{Fraction}
\end{eqnarray}
where $F_{\rm ads}$ and $F_{\rm tail}$ are    free energies  of the adsorbed and
desorbed  portions of the chain respectively. The free energy  $F_{\rm ads} =
k_BT [ N - M(h, \epsilon)] \ln z^{*}(\epsilon)$ whereas $F_{\rm tail} = k_BT
\mu_{\rm des} M(h, \epsilon)$ (recall that $\mu_{\rm des}$ is the chemical
potential of a desorbed monomer). After substitution of  these expressions in eq.
(\ref{Fraction}) and taking into account that in equilibrium $\mu_{\rm ads} =
\mu_{\rm des}$ (the sequence of operations is important: taking the derivative
with respect to $\epsilon$ is to be followed by the condition $\mu_{\rm ads}
=\mu_{\rm des}$) so one gets
\begin{eqnarray}
 n =  - \left[1 - \frac{M(h, \epsilon)}{N}\right] \frac{\partial \ln
z^{*}(\epsilon)}{\partial
\epsilon}
\label{Order_Parameter}
\end{eqnarray}
i.e. the order parameter $n$ is defined by the product of monomer fraction in
the adsorbed portion, $1 - M/N$, and the fraction of surface contacts in this
portion, $ - \partial \ln z^{*}/\partial \epsilon$.
 The expressions for the order parameter can be recast in the form
\begin{eqnarray}
 n = \left|\frac{\partial \ln
z^{*}(\epsilon)}{\partial
\epsilon}\right| \times \begin{cases}
                         1 - \frac{h}{c_1 l_0 N} \left(\frac{k_BT}{l_0 f_{\rm
p}}\right)^{1/\nu - 1} &\mbox{,   BS-model}\\
\\
1 - \frac{h}{c_2 l_0 N} \left[{\cal L} \left(\frac{l_0 f_{\rm
p}}{k_BT}\right)\right]^{-1}
&\mbox{,   FJVB-model}
\end{cases}
\label{Order_Parameter1}
\end{eqnarray}
 Here $c_1$ and $c_2$ are some  constants of the order of unity.

As one can see from Eq.~(\ref{Order_Parameter1}), the order parameter decreases
linearly and steadily with $h/N$. This behavior is qualitatively different from
the abrupt jump of $n$ when the pulling force $f$ is changed as a control
parameter. In Section \ref{MC_results} we will show that this predictions is in
a good agreement with our MC - findings. The transition point on the $n$ vs. $h$
 curve corresponds to total detachment, $n = 0$.  The corresponding distance $h$
will be termed ``detachment height'' $h_D$. The dependence of $h_D$ on the
adsorption energy $\epsilon$ can be obtained from Eq.(\ref{Order_Parameter1})
where $n$ is set to zero, i.e.
\begin{eqnarray}
\frac{h_D}{l_0 N}  = \begin{cases}
   \left( \frac{l_0 f_{\rm p}}{k_BT}\right)^{1/\nu -1}  &\mbox{, BS-model}\\
\\
{\cal L}\left( \frac{l_0 f_{\rm p}}{k_BT}\right) &\mbox{,  FJVB-model}
                     \end{cases}
\label{Detachment_Line}
\end{eqnarray}
where again $f_{\rm p}$ as a function of $\epsilon$ is given by Eq.
(\ref{Local_Equilibrium}).
The line given by Eq. (\ref{Detachment_Line}), is named ``detachment line''. It
corresponds to an adsorption - desorption polymer transition which appears as of
{\it second order} since this order parameter $n$ goes to zero continuously as
$h$ increases.  One should emphasize, however, that this ``detachment''
transition has the same nature as the force-induced desorption transition
\cite{Bhattacharya} in the $f$-ensemble where the pulling force $f$, rather than
the distance $h$, is fixed and used as a control parameter. This phase
transformation is known to be of {\it first order}.

It is easy to understand (cf. with Eq.(\ref{M_vs_h}) ) that the condition $M(h,
\varepsilon) = N$ corresponds to the detachment line as well as
to a terminal point of the force  plateau. It can be seen in the MC -
simulation results in Section\ref{MC_results}, Fig.~\ref{force}.

\section{Probability distribution $P(K)$ of the number of adsorbed
monomers}\label{Theory2}

The grand canonical ensemble (GCE) method, which has been used in our recent
paper \cite{Bhattacharya}, is a good starting point to calculate the probability
distribution function $P(K)$ of  the adsorbed monomers number $K$. According to
this approach, the GCE-partition function of an adsorbed chain has the form
\begin{equation}
\Xi (z, w) = \sum_{N=1}^{\infty} \: \sum_{K=0}^{\infty} \Xi_{N, K} \: z^N \:
w^K = \dfrac{V_0 (wz) \: Q (z)}{1 - V(w z)\:  U(z)}
\label{GCE}
\end{equation}
where $z$ and $w = \exp(\epsilon)$ are the fugacities conjugated to chain length
$N$ and to the number of adsorbed monomers $K$, respectively. In Eq.(\ref{GCE})
$U(z)$, $V(wz)$ and $Q(z)$ denote the GCE partition functions for  loops, trains
and tails, respectively. The building block adjacent to the tethered chain end
corresponds to $V_0 (wz) = 1 + V (wz)$. It has been shown \cite{Bhattacharya}
that the functions $U(z)$, $V(wz)$ and $Q(z)$ can be expressed in terms of
polylog functions, defined in the paragraph after Eq. (\ref{Basic}), as $U (z) =
\Phi (\alpha, \mu_3 z)$, $V (wz) = \Phi (1 - \gamma_{d=2}, \mu_2 w z)$ and $Q
(z) = 1 + \Phi (1 - \gamma_{1}, \mu_3  z)$, where $\mu_3$ and $\mu_2$ are $3d$-
and $2d$ - connective constants respectively. By making use of the inverse
Laplace transformation of $\Xi (z, w)$ with respect to  $z$ (see, e.g.
\cite{Rudnick}) the (canonical with respect to the chain length $N$)  partition
function is obtained as
\begin{equation}
\Xi_{N}(w) = \sum_{K=0}^{\infty} \: \Xi_{N,K} \: {w}^{K} =\exp[-N
\ln z^{*} (w)]
\label{GC}
\end{equation}
where $z^{*}(w)$ is a simple pole of $\Xi (z, w)$ in complex $z$-plane given by
equation $V(w z^*) U(z^*) = 1$, i.e. by Eq.(\ref{Basic}).

The (non-normalized) probability for the chain to have $K$ adsorbed monomers is
$P(K) \propto \Xi_{N,K} \: {\rm e}^{\epsilon K} = \exp[- {\cal F}(K)/k_BT]$,
where ${\cal F}(K)$ is the free energy at given $K$. It is convenient to
redefine the fugacity $w$ as $w \rightarrow \xi  w$, (as well as $w_c
\rightarrow \xi  w_c$) where $\xi$ is an arbitrary complex variable. Then the
probability $P(K)$ can be found as the coefficient of $\xi^K$ in the function
$\Xi_{N} (\xi w)$ expansion in powers of $\xi$. Therefore
\begin{equation}
 P(K) = \exp[- {\cal F}(K)/k_BT] = \frac{1}{2\pi i} \: \displaystyle \oint
\dfrac{\exp[- N \ln z^{*} (\xi \: w)]}{\xi^{K+1}} \: d \xi
\label{Integral}
\end{equation}
where the contour of integration is a closed path in the complex $\xi$ plane
around $\xi =0$. (see e.g. \cite{Rudnick}). To estimate  the integral in
Eq.(\ref{Integral}) we use the {\it steepest descent method} \cite{Rudnick}.

For large $N$ the main contribution to the integral in Eq. (\ref{Integral}) is
given by the saddle point $\xi = \xi_0$ of the integrand which is defined by the
extremum of the function $g(\xi) = - \ln z^{*}(\xi \: w) -
[(K+1)/N] \ln \xi$, i.e., by the condition
\begin{equation}
 \dfrac{K+1}{\xi_0} = - \left. N \: \dfrac{\partial \ln z^{*}}{\partial
\xi}\right|_{\xi = \xi_0}
\label{SP}
\end{equation}
The integral is dominated by the term $\exp [- N \ln z^{*}(\xi_0 w) - (K+1)
\ln \xi_0]$. Another contribution comes from  the
integration along the steepest descent line. As a  result one obtains
\begin{eqnarray}
 P(K) \propto  \dfrac{\exp [- N \ln z^{*}(\xi_0 \: w) - (K+1) \ln
\xi_0]}{\sqrt{N\left[\left(\dfrac{N}{K}\right)
\left[\dfrac{\partial\ln z^{*}}{\partial \xi}\right]_{\xi = \xi_0}^2 -
\left.\dfrac{\partial^2 \ln z^{*}}{\partial \xi^2}\right|_{\xi = \xi_0}\right]}}
\label{Method}
\end{eqnarray}
The validity of the steepest descent method is ensured by the condition of
the second derivative $N g''(\xi_0)$ being large, which yields
\begin{eqnarray}
 N g''(\xi_0) = N\left[\left(\dfrac{N}{K}\right)
\left[\dfrac{\partial\ln z^{*}}{\partial \xi}\right]_{\xi = \xi_0}^2 -
\left.\dfrac{\partial^2 \ln z^{*}}{\partial \xi^2}\right|_{\xi = \xi_0}\right]
\gg 1
\label{Valid}
\end{eqnarray}
A more explicit calculation whithin this method can be performed in the
vicinity of the critical point $\epsilon = \epsilon_c$.

\subsection{PDF of the number of adsorbed monomers close to the critical point
of adsorption}

In this case the explicit form of $\ln z^{*}$ is known \cite{Bhattacharya} and
after the redefinition of the fugacity, $w \rightarrow \xi w$, it reads
\begin{eqnarray}
 \ln z^{*} (\xi w) = - a_1 (w - w_c)^{1/\phi} \xi^{1/\phi} - \ln \mu_3
\label{Z_Star}
\end{eqnarray}
where $a_1$ is a constant of the order of unity. The critical adsorption
fugacity $w_c = {\exp (\epsilon_c)}$ is defined by the equation
\begin{eqnarray}
 \zeta (\alpha)\: \Phi (1 - \gamma_{d=2}, \mu_2 w_c/\mu_3) = 1
\label{Critical}
\end{eqnarray}
with $\zeta (x)$ denoting the Riemann zeta-function.

By using Eq.(\ref{Z_Star}) in Eq.(\ref{SP}) one arrives at the expression for
the saddle point
\begin{eqnarray}
 \xi_0 = \left( \dfrac{K}{N}\right)^{\phi} \dfrac{a_2}{w - w_c}
\label{SP_Solution}
\end{eqnarray}
where $a_2=(\phi/a_1)^{\phi}$. Using Eq. (\ref{Z_Star}) and Eq.
(\ref{SP_Solution}) in
Eq. (\ref{Method}) then yields the expression for PDF
\begin{eqnarray}
 P(K) \propto [(w - w_c) {\rm e}^{a_2}]^{K}  \left(\dfrac{K}{N}\right)^{- \phi K
- 1/2} \: \dfrac{\mu_{3}^{N}}{N^{1/2}}
\label{Distr1}
\end{eqnarray}
For reasonably large $K$ and after normalization one arrives at the final
expression for the PDF
\begin{eqnarray}
  P(K) = \dfrac{\eta^K}{{\cal C}(\eta) K^{\phi K}  }
\label{Distr3}
\end{eqnarray}
where we have introduced the usual adsorption scaling variable $\eta = b_1 (w -
w_c)N^\phi$ ($b_1$ is a constant of the order of unity; see e.g. \cite{Hsu}) as
well as the normalization constant
${\cal C}$:
\begin{eqnarray}
 {\cal C}(\eta) = \sum_{K=1}^{N} \: \dfrac{\eta^K}{K^{\phi K}}
\label{Normalization}
\end{eqnarray}
One can readily see that the width of the distribution increases with $w$
or with $\epsilon$. To this end one may directly calculate the fluctuation
variance as follows:
\begin{eqnarray}
 \overline{(K - \overline{K})^2} = - N  \dfrac{\partial^2 \ln z^{*}(w)}{\partial
(\ln w )^2} \propto  N w (w/\phi - w_c) (w - w_c)^{1/\phi - 2}
\label{Variance_1}
\end{eqnarray}
where the expression for $\ln z^{*}(w)$ given by Eq.(\ref{Z_Star})
(where also $\xi=1$ ) has been used. Taking into account that $\phi \approx
0.5$, it becomes clear that the variance really grows with $w$.

The validity of the steepest descent method is ensured by the condition
Eq.(\ref{Valid}). Using Eq.(\ref{Z_Star}) and Eq. (\ref{SP_Solution}), one can
verify that this criterion holds when $N (w - w_c)^2 (K/N)^{1 - 2\phi} \gg 1$.
We recall that $\phi \approx 0.5$, so that $(w - w_c) N^{1/2} \gg 1$.  In
result the criterion becomes
\begin{eqnarray}
 N^{-1/2} \ll  (w - w_c) \ll 1
\label{Criterion}
\end{eqnarray}
In the deep adsorption regime this condition might be violated. Nevertheless,
the steepest descent method could still be used there, provided that the
appropriate solution for $z^{*}(w)$ (see Eq.(\ref{Basic})) is chosen.

\subsection{The regime of deep adsorption}

In the deep adsorption regime one should use the the solution for $z^{*} (w)$
which was also discussed in ref. \cite{Bhattacharya}. Namely,  in this case
\begin{eqnarray}
 z^{*} (w) \approx \dfrac{1}{\mu_2 w} \left[1 - \left( \dfrac{\mu_3}{\mu_2
w}\right)^{1/(1-\lambda)} \right]
\label{Deep_Ads}
\end{eqnarray}
With Eq.~(\ref{Deep_Ads}) the mean value $\overline{K} = - N \partial \ln
z^{*}(w)/\partial \ln w$ can be written as
 \begin{eqnarray}
\dfrac{\overline{K}}{N} = 1 - \dfrac{1}{1 - \lambda} \left(
\dfrac{\mu_3}{\mu_2 w}\right)^{1/(1-\lambda)}
\label{K}
 \end{eqnarray}
thus $\overline{K}$ tends to $N$  with $w$ growing as it should be. The
variance of the fluctuations within the GCE then becomes
 \begin{eqnarray}
\overline{(K - \overline{K})^2} = - N  \dfrac{\partial^2 \ln
z^{*}(w)}{\partial (\ln w )^2} = \dfrac{N}{(1 - \lambda)^2}
\left(\dfrac{\mu_3}{\mu_2 w}\right)^{1/(1-\lambda)},
\label{Variance_2}
 \end{eqnarray}
i.e., the fluctuations decrease when the adsorption energy grows. Comparison
of this result with the result given by Eq. (\ref{Variance_1}) leads to the
important conclusion that the fluctuations of the number of adsorbed monomers
first grow with $\epsilon$, attain a maximum, and finally decrease with
increasing surface adhesion $\epsilon$. The position of the maximum reflects the
presence of finite-size effects, and, as the chain length $N\to \infty$, this
maximum occurs at the CAP.

Consider now the steepest descent treatment for the deep adsorption regime. With
Eq.(\ref{Deep_Ads}) (after the rescaling $w \rightarrow \xi w$) in
Eq.(\ref{SP}), the saddle point becomes
 \begin{eqnarray}
\xi_0 = b_2 \left(\dfrac{\mu_3}{\mu_2 w} \right) \left(1 -
\dfrac{K}{N}\right)^{-(1 - \lambda)}
\label{SP_Deep}
 \end{eqnarray}
where $b_2 = (1 - \lambda)^{-(1 - \lambda)}$ (recall that $1-\lambda =
\gamma_{d=2} = 1.343$).  The main contribution comes from the exponential term
in Eq.(\ref{Method}) which is given  by
\begin{eqnarray}
 \exp\left[-N \ln z^{*} (w \xi_0) - (K+1) \ln \xi_0\right] = (\mu_2 w)^{N}
\dfrac{\left(b_2 \mu_3 {\rm e}^{1-\lambda} N^{1-\lambda}/\mu_2
w\right)^{N-K}}{\left(N - K\right)^{(1-\lambda)(N-K)}}
\label{Exponential}
\end{eqnarray}
For $N-K \gg 1$  the expression for the non-normalized PDF takes on the form
\begin{eqnarray}
P(K) \propto \dfrac{\left(b_3 \mu_3  N^{1-\lambda}/\mu_2
w\right)^{N-K} }{(N-K)^{(1-\lambda) (N - K)} }
\end{eqnarray}
where $b_3$ is a constant of the order of unity. Normalization of this
distribution yields
\begin{eqnarray}
 P(K) = \dfrac{\chi^{N-K}
}{{\cal R}(\chi) (N-K)^{(1-\lambda)(N - K)} }
\label{PDF_Deep}
\end{eqnarray}
where the parameter $\chi = b_3 (\mu_3/\mu_2 w) N^{1-\lambda}$ and the
normalization
constant
\begin{eqnarray}
 {\cal R}(\chi) = \sum_{K=1}^{N-1} \dfrac{\chi^{N-K}
}{(N-K)^{(1-\lambda)(N - K)} }.
\label{R}
\end{eqnarray}
The validity condition, Eq.(\ref{Valid}), in the deep adsorption regime
(after substitution of Eq. (\ref{Deep_Ads}) into Eq.(\ref{Valid})) requires
\begin{eqnarray}
 (N - K)^{3/2 - \lambda} \gg (\mu_3/\mu_2 w) N^{1 - \lambda},
\label{Valid_Deep}
\end{eqnarray}
i.e., $K$ should not be very close to $N$. In Fig.~\ref{pdf_theory}a we show the
PDF of the number of chain contacts, $P(K)$, for a free chain without pulling
and several adsorption strengths of the substrate. One can readily verify that
visually the shape of $P(K)$ resembles very much a Gaussian distribution at
moderate values of $\epsilon_c\approx 1.7 < \epsilon < 6.0 $. The PDF
variance goes through a sharp maximum at $\epsilon \gtrsim \epsilon_c$ and then
declines, as expected from Eq.~(\ref{Variance_2}).

One should note that in the $f$-ensemble (where the force $f$ and not the
distance $h$ acts as a controll parameter \cite{Bhattacharya}) the order
parameter $n = \overline{K}/N$ undergoes a jump at the detachment adsorption
energy $\epsilon_D$. This means that $N (\partial^2 \ln z^{*}/\partial
\epsilon^2)_{\epsilon_D} = - (\partial \overline{K}/\partial
\epsilon)_{\epsilon_D} \rightarrow - \infty$. Thus at the detachment point the
variance of the fluctuations $\overline{(K - \overline{K})^2} = - N (\partial^2
\ln z^{*}/\partial \epsilon^2)_{\epsilon_D} \rightarrow \infty$, which
practically means that for chains of a finite length the distribution at
$\epsilon = \epsilon_D$ becomes very broad, in sharp contrast to
Eq.~(\ref{Variance_2}). This has indeed been observed in our MC-simulation
results (see Fig.12 in ref. \cite{Bhattacharya}).

\subsection{$P(K)$ distribution in the subcritical regime $w < w_c$ of
underadsorption}

In the subcritical regime, $w < w_c$, the fraction of adsorbed points (order
parameter) $n = \overline{K}/N = 0$, in the thermodynamic limit. Nevertheless,
$\overline{K} \neq 0$ and one can examine the form of the PDF $P(K)$. At $w <
w_c$ the solution for $z^* (w)$ (the simple pole of $\Xi (z, w)$ in the complex
$z$-plane) does not exist because $V(w z) U(z) < 1$ (see Eq.(\ref{GCE})).
However, the tail GCE-partition function  $Q(z) = 1 + \Phi (1- \gamma_1, \mu_3
z) \propto \Gamma (\gamma_1)/(1 - \mu_3 z)^{\gamma_1}$ has a {\em branch point}
at $z = 1/\mu_3$ (see Eq. (A 11) in ref. \cite{Bhattacharya}) which governs the
coefficient at $z^N$, i.e., the partition function $\Xi_{N}(w)$. The calculation
(following Section 2.4.3 in ref. \cite{Rudnick}) yields
\begin{eqnarray}
 \Xi_{N}(w) = \dfrac{1 + \Phi (\lambda, \mu_2 w/\mu_3)}{1 - \zeta
(\alpha) \: \Phi (\lambda, \mu_2 w/\mu_3)} \: \mu_3^N \: N^{\gamma_{1}
- 1}
\label{Subcritical_GC}
\end{eqnarray}
where $\lambda = 1 - \gamma_{d=2}$ and we have also used that at $z = 1/\mu_3$
the loop and train GCE-partition
functions are $U(1/\mu_3)= \Phi (\alpha, 1) = \zeta
(\alpha)$ and $V(w/\mu_3) = \Phi (\lambda, \mu_2 w/\mu_3)$
respectively.
This expression has a pole at $w = w_c$ (cf. Eq. (\ref{Critical})) which yields
the coefficient of $w^K$ , i.e. $\Xi_{N,K}$. Recall that $\Xi_N(w) =
\sum_{K=0}^{\infty} \Xi_{N,K} w^K$, so that $P(K) \propto \Xi_{N,K} w^K$.
Expansion of the denominator in Eq. (\ref{Subcritical_GC}) around $w = w_c$
reveals the simple pole as follows
\begin{eqnarray}
 \Xi_{N}(w) = \dfrac{[1 + 1/\zeta(\alpha)] \: \mu_3^N  N^{\gamma_1 -
1}}{\zeta(\alpha) \: \Phi (-\gamma_{d=2}, \mu_2 w_c/\mu_3)} \left(
\dfrac{w_c}{w_c - w}\right).
\label{Pole}
\end{eqnarray}
In (\ref{Pole}) we have used the relationship $x (d/d x) \Phi (1-\gamma_{d=2},
x) = \Phi(- \gamma_{d=2}, x)$. The coefficient of $w^K$, i.e., $\Xi_{N,K}$, is
proportional to $w_c^{- K}$. Therefore $P(K) \propto (w/w_c)^{K} = \exp
[-(\epsilon_c - \epsilon)K]$. Taking the  normalization condition
$\sum_{K=0}^{N} P(K) = 1$ into account, the final expression for $P(K)$ can be
recast in the form
\begin{eqnarray}
 P(K) = \dfrac{1 - {\exp}[-(\epsilon_c - \epsilon)]}{1 - {\exp}[- N
(\epsilon_c - \epsilon)]} \: {\exp}[- K(\epsilon_c - \epsilon)],
\label{Exp}
\end{eqnarray}
i.e., $P(K)$ has  a simple exponential form. The calculation of the average
$\overline{K} = \sum_{K=0}^{N} K P(K)$ leads to the simple result $\overline{K}
= [\exp (\epsilon_c - \epsilon) - 1]^{-1} \approx 1/(\epsilon_c - \epsilon)$,
i.e., $\overline{K} \rightarrow \infty$ at $\epsilon \rightarrow \epsilon_c$. On
the other hand we know that  $\overline{K} = N^{\phi}$ at $\epsilon =
\epsilon_c$. In order to prevent a divergency at $\epsilon = \epsilon_c$, one
should incorporate an appropriate cutoff in the PDF given by Eq.(\ref{Exp}).
With this the distribution is given by
\begin{eqnarray}
 P(K) = \dfrac{1 - {\exp}[-(\epsilon_c - \epsilon + 1/N^{\phi})]}{1 - {\exp}[- N
(\epsilon_c - \epsilon + 1/N^{\phi})]} \: {\exp}[- K(\epsilon_c - \epsilon +
1/N^{\phi})].
\label{Exp_Cutoff}
\end{eqnarray}
Thus the expression for the average number of adsorbed monomer has the correct
limit behavior, i.e.,
\begin{eqnarray}
 \overline{K} = \dfrac{1}{\epsilon_c - \epsilon + 1/N^{\phi}}
\label{Average_Cutoff}
\end{eqnarray}

\subsection{Probability distribution function $P(K)$ in the $h$-ensemble}

Eventually we examine how the fixed chain-end hight $h$ affects the PDF of the
number of contacts $K$. To this end we refer again to Fig. \ref{Tearing}a where
an adsorbed chain with a fixed height $h$ of the last monomer is depicted. The
adsorbed chain consists of a tail of length $M$ and of an adsorbed part with $N
- M$ beads. One should bear in mind that $M$ is function of the control
parameters $h$ and of $w = \exp (\epsilon)$ given by
eqs.(\ref{Local_Equilibrium})
and (\ref{M_vs_h}).
The partition function of the adsorbed part is then given by
\begin{eqnarray}
 \Xi_{N} (w) = \sum_{K=0}^{N} \: \Xi_{N, K} \: w^K  = \exp\left\lbrace - [N -
M(h, w)] \ln z^{*} (w)\right\rbrace
\label{Adsorbed_Part}
\end{eqnarray}
where we took into account that the free energy of the adsorbed portion is given
as $F_{\rm ads} = k_B T [N - M(h, w)] \ln z^{*}(w)$ (see Sec. II B).

\begin{figure}[bht]
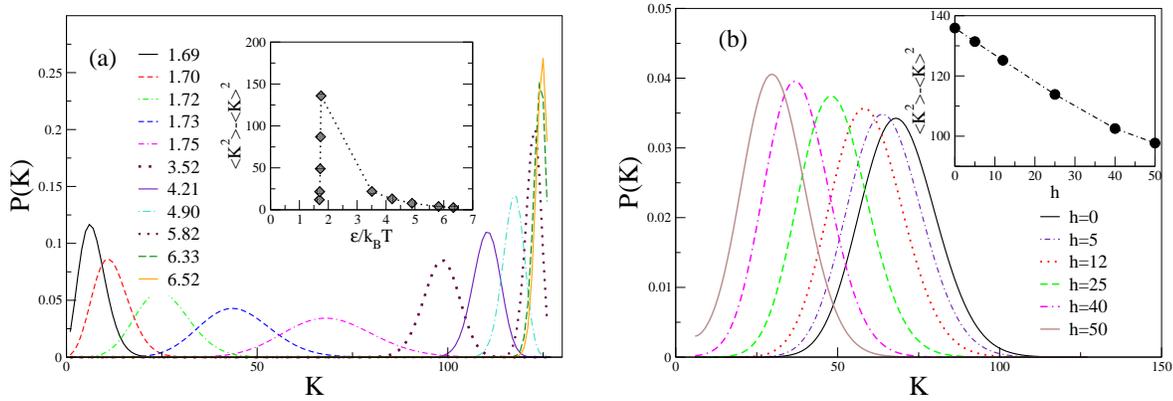

\hspace{-2.0cm}
\includegraphics[scale=0.3, angle=0]{PDF_theor.eps}
\hspace{0.50cm}
\includegraphics[scale=0.3, angle=0]{PDF_pull.eps}
\caption{\label{pdf_theory} (a) Probability Distribution Function of the number
of chain contacts with the grafting plane,  $K$, for different degrees of
adhesion $\epsilon$ and no pulling force $f = 0$ as  follows from eqs.
(\ref{Distr3}) and (\ref{PDF_Deep}). The variance of $P(K)$ is
shown in the inset for different $\epsilon$. (b) The same as in (a) for
$\epsilon = 1.75$ and different heights of the chain end $h$. The change of the
variance $\langle K^2\rangle - \langle K\rangle^2$ with $h$ is shown in the
inset.}
\end{figure}
As mentioned above, the PDF $P(K) \propto \Xi_{N, K} \: w^K $, so that by
means of rescaling $w \rightarrow \xi w$ and $w_c \rightarrow \xi w_c$  the PDF
can be found as the coefficient of $\xi^K$, i.e.,
\begin{equation}
 P(K)  = \frac{1}{2\pi i} \: \displaystyle \oint
\dfrac{\exp\left\lbrace - [N -
M(h, \xi w)] \ln z^{*} (\xi w)\right\rbrace}{\xi^{K+1}} \: d
\xi
\label{Coefficient}
\end{equation}
As before, the steepest descent method can be used to calculate the
integral in Eq. (\ref{Coefficient}). However, in this case the calculations are
more complicated and we have relegated most of them to Appendix B. As may  be
seen there, the saddle point equation can not be solved analytically in
the general case but could be treated iteratively for relatively small heights
$h$.

One can readily see that for $h = 0$ Eq. (\ref{PDF_Pull}) reduces to
Eq.(\ref{Distr3}). The PDF, following from Eq. (\ref{PDF_Pull}) is shown in
Fig. \ref{pdf_theory}b for $\epsilon = 1.75$ and several values of the height
$h$. It can be seen that the curve for $h = 0$ coincides with the curve for
$\epsilon = 1.75$ in Fig. \ref{pdf_theory}a as it should be. Evidently, both
the mean value $\overline{K}$ and $\overline{K^2} - \overline{K}^2$ decline
with growing $h$.

\section{Monte Carlo Simulation Model}\label{MC_model}

We use a coarse grained off-lattice bead-spring model \cite{MC_Milchev} which
has proved rather efficient in a number of polymers studies so far. The system
consists of a single polymer chain tethered at one end to a flat impenetrable
structureless surface. The surface interaction is described by a square well
potential,
 \begin{equation}\label{ads_pot}
 U_w(z) = \begin{cases}
\epsilon, & z < r_c \\
0, & z \ge r_c
          \end{cases}
\end{equation}
The strength $\epsilon$ is varied from $2.0$ to $5.0$ while the
interaction range $r_c = 0.125$. The effective bonded interaction is described
by the FENE (finitely extensible nonlinear elastic) potential:
\begin{equation}
U_{FENE}= -K(1-l_0)^2ln\left[1-\left(\frac{l-l_0}{l_{max}-l_0} \right)^2 \right]
\label{fene}
\end{equation}
with $K=20$, and the fully stretched-, mean-, and minimum bond lengths
$l_{max}=1, l_0 =0.7, l_{min} =0.4$. The nonbonded interactions between monomers
are described by the Morse potential:
\begin{equation}\label{Morse}
\frac{U_M(r)}{\epsilon_M} =\exp(-2\alpha(r-r_{min}))-2\exp(-\alpha(r-r_{min}))
\end{equation}
with $\alpha =24,\; r_{min}=0.8,\; \epsilon_M/k_BT=1$. In few cases, needed to
clarify the nature of the polymer chain resistance to stretching, we have taken
the nonbonded interactions between monomers as purely repulsive by shifting the
Morse potential upward by $\epsilon_M$ and removing its attractive branch for
$r \ge r_{min}$.

We employ periodic boundary conditions in the $x-y$ directions and impenetrable
walls in the $z$ direction. The lengths of the studied polymer chains are
typically $64$, and  $128$. The size of the simulation box was chosen
appropriately to the chain length, so for example, for a chain length of $128$,
the box size was $256 \times 256 \times 256$ . All simulations were carried out
for constant position of the last monomer $z$-coordinate, that is, in the fixed
height ensemble. The the fluctuating force $f$, exerted on the last bead by the
rest of the chain was measured and average over about $2000$ measurements.
The standard Metropolis algorithm was employed to govern the moves with  self
avoidance automatically incorporated in the potentials. In each Monte Carlo
update, a monomer was chosen at random and a random displacement attempted with
$\Delta x,\;\Delta y,\;\Delta z$ chosen uniformly from the interval $-0.5\le
\Delta x,\Delta y,\Delta z\le 0.5$. If the last monomer was displaced in $z$
direction, there was an energy cost of $-f\Delta z$ due to the pulling force.
The transition probability for the attempted move was calculated from the change
$\Delta U$ of the potential energies before and after the move was performed as
$W=exp(-\Delta U/k_BT)$. As in a standard Metropolis algorithm, the attempted
move was accepted, if $W$ exceeds a random number uniformly distributed in the
interval $[0,1)$.

As a rule, the polymer chains have been originally equilibrated in the MC method
for a period of about $ 5 \times 10^5$ MCS after which typically $500$
measurement runs were performed, each of length $2 \times 10^6$ MCS. The
equilibration period and the length of the run were chosen according to the
chain length and the values provided here are for the longest chain length.

\section{Monte Carlo simulation results}\label{MC_results}

In order to verify the theoretical predictions, outlined in Section
\ref{Theory}, we carried out extensive Monte Carlo simulations with the
off-lattice model, defined in Section \ref{MC_model}. In these simulations we
fix the end monomer of the polymer chain at height $h$ above the adsorbing
surface, and measure the (fluctuating) force, needed to keep the last bead at
distance $h$, as well as the corresponding fraction of adsorbed monomers $n$.
These computer experiments are performed at different strengths $\epsilon$ of
the adsorption potential, Eq.~(\ref{ads_pot}).
\vspace{0.5cm}
\begin{figure}[bht]
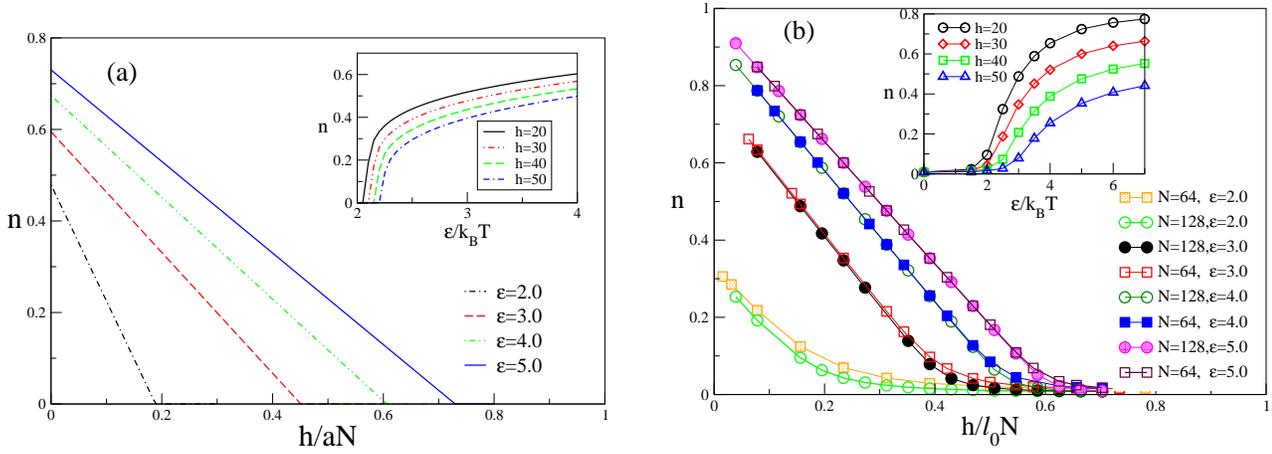

\hspace{-2.0cm}
\includegraphics[scale=0.33]{op_theor.eps}
\hspace{0.50cm}
\includegraphics[scale=0.34]{op_MC.eps}
\caption{\label{OP} (a) Order parameter (fraction of adsorbed monomers) $n$
variation with changing height $h/l_0N$ of the fixed chain-end for polymers of
length $N = 128$ and different adsorption strength $\varepsilon/k_BT$. (b)
Variation of $n$ with $\varepsilon/k_BT$ for different fixed positions of the
chain-end $h/l_0N$ as it is seen from MC-data. Insets show the resulting $n-
\epsilon$ relationship at
several fixed heights $h$.}
\end{figure}
In Figs.~\ref{OP}a,b we compare the predicted dependence of the order parameter
$n$ on the (dimensionless) height $h/l_0N$ at several values of $2.0 \le
\epsilon \le 5.0$ with the results from MC simulations. Note, that the critical
point of adsorption $\epsilon_c \approx 1.69$ so we take our measurements above
the region of critical adsorption. Typically, both in the analytic results,
Fig.~\ref{OP}a, and
 in the MC-data, Fig.~\ref{OP}b, for $N=128$, one recovers the predicted linear
decrease of $n$ with growing $h$. Finite-size effects lead to some rounding of
the simulation data (in Fig.~\ref{OP}b these effects are seen to be larger for
$N=64$ than for $N=128$) when $n\to 0$ so that the height of detachment $h_D$ is
determined from the intersection of the tangent to $n(h)$ and the $x-$axis where
\begin{figure}[htb]
 \includegraphics[scale=0.4]{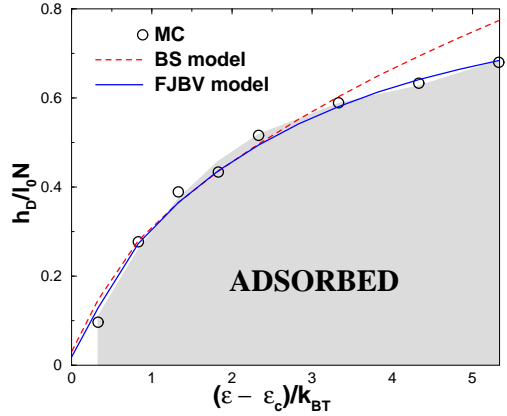}
\caption{Phase diagram showing the dependence of the critical height of polymer
detachment from the substrate, $h_D/l_0 N$, with the relative strength of
adsorption $(\varepsilon - \varepsilon_{c})/k_BT $, where $\varepsilon_{c}/k_BT$
is the critical point of adsorption  at  zero force. The theoretical curves
follow Eq. (\ref{Detachment_Line}).}
\label{phase_diag}
\end{figure}
$n=0$. Evidently, with growing adsorption strength, $\epsilon$, larger height
$h_D$ is needed to detach the polymer from the substrate. Thus, one may
construct a phase diagram for the desorption transition, which we show in
Fig.~\ref{phase_diag}. The theoretical prediction is given by eq.
(\ref{Detachment_Line}).

In the insets of Fig.~\ref{OP}a,b we also show the variation of the fraction of
adsorbed segments $n$ with adsorption strengths $\epsilon$ for several heights
$20 \le h \le 50$ of a chain with $N = 128$. It is evident that, apart from the
rounding of the MC data at $n \to 0$, one finds again good agreement between the
behavior, predicted by Eq.~(\ref{Order_Parameter}), and the simulation results.

The gradual change of $n$ in the whole interval of possible variation of $h$
suggests a pseudo-continuous phase transition, as pointed out in the end of
Section \ref{Theory}, Eq.~(\ref{Order_Parameter1}). Of course, if $h$ is itself
expressed in terms of the measured pulling force, one would again find that $n$
changes abruptly with varying $f$ at some threshold value $f_D$, indicating a
first order transformation from an adsorbed into a desorbed state of the polymer
chain.
\vspace{0.9cm}
\begin{figure}[htb]
 \includegraphics[scale=0.4]{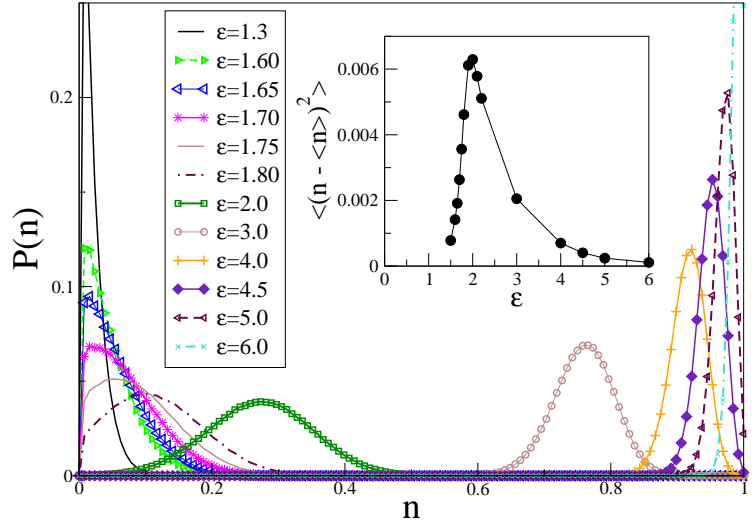}
\caption{PDF of the order parameter (fraction of contacts with the plane) for
different adsorption strength $\epsilon$ at zero force. The critical adsorption
point (CAP)
$\epsilon_{c} \approx 1.67$. The change of the variance $\langle n^2\rangle -
\langle n\rangle^2$ with varying $\epsilon$ is displayed in the inset.
}
\label{pdf_MC}
\end{figure}

It has been pointed out earlier by Skvortsov et al.\cite{Skvortsov} that, while
both the fixed-force and the fixed-height ensembles are equivalent as far as the
mean values of observables such as the fraction of adsorbed monomers and other
related quantities are concerned, this does not apply to some more detailed
properties like those involving fluctuations. Therefore, it is interesting to
examine the fluctuations of the order parameter, $n$, for different values of
our control parameter $h$, and compare them to theoretical predictions for
$P(K)$ from Section \ref{Theory}. First we compare the order parameter
distribution $P(n)$ for zero force, Fig.~\ref{pdf_MC}, obtained from our
computer experiment, to that, predicted by Eqs.~(\ref{Distr3}),
(\ref{PDF_Deep}), (\ref{Exp_Cutoff}), and displayed in Fig.~\ref{pdf_theory}a.
It is evident from Fig.~\ref{pdf_MC} that for free chains at different strengths
of adhesion there is a perfect agreement between analytical and simulational
results. For rather weak adsorption $\epsilon = 1.3\pm 1.6 < \epsilon_c=1.67$ in
the subcritical regime, one can verify from Fig.~\ref{pdf_MC} that $P(n)$
gradually transforms from nearly Gaussian into exponential distribution, as
expected from Eq.~(\ref{Exp_Cutoff}). For $\epsilon > \epsilon_c$ the
distribution width $\langle (n - \langle n \rangle)^2\rangle$ grows and goes
through a sharp maximum in the vicinity of $\epsilon_c$, and then drops as
$\epsilon$ increases further - compare insets in Fig.~\ref{pdf_MC} and
Fig.~\ref{pdf_theory}a.
\vspace{0.6cm}
\begin{figure}[htb]
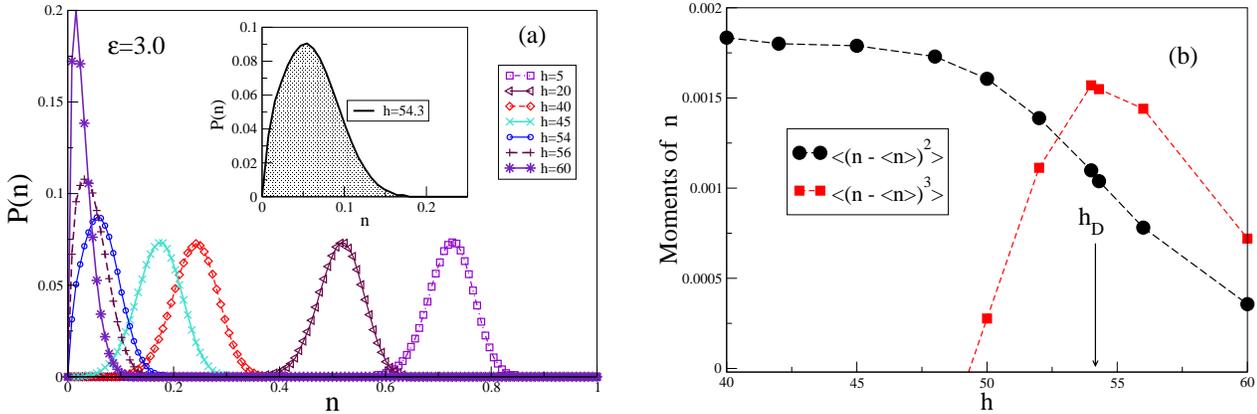

\hspace{-2.0cm}
\includegraphics[scale=0.33]{ophistogram.eps}
\hspace{0.50cm}
\includegraphics[scale=0.33]{moments.eps}
\caption{\label{PDF_OP} (a) Probability distribution $P(n)$ of the order
parameter $n$ (i.e., the fraction of adsorbed monomers) for $N = 128$ and
$\epsilon = 3.0$ at different heights of the chain-end $h$ over the grafting
plane. In the inset we show $P(n)$ at the detachment line $h_D = 54.3$. (b)
Variation of the second- and third central moments of $P(n)$ with $h$. The
maximum of $\langle (n - \langle n \rangle)^3\rangle$ is reached at $h = h_D$.}
\end{figure}

Let us consider now PDF in the presence of pulling.
In Fig.~\ref{PDF_OP}a we display the distribution $P(n)$ measured in the MC
simulations for different heights $h$ and constant adsorption energy $\epsilon =
3.0$. One can readily verify from our results that far enough from the
detachment line, $h < h_D$, the shape of $P(n)$ looks like Gaussian and that the
second moment, $\langle (\Delta n)^2\rangle$, remains unchanged with varying
height $h$. Of course, when $h \to h_D$, the maximum of $P(n)$ shifts to lower
values of $n$. Only in the immediate vicinity of $h_D$, where $n \to 0$ and the
fluctuations strongly decrease, one observes a significant deviation from the
Gaussian shape - cf. the inset in Fig.~\ref{PDF_OP}a. The latter is illustrated
in more detail in Fig.~\ref{PDF_OP}b where we show the measured variation of the
second moment, $\langle (\Delta n)^2\rangle$, and that of the third moment,
$\langle (n - \langle n \rangle)^3\rangle$ with increasing height $h$. The
deviation from Guassinity in $P(n)$, measured by the
deviation of the third moment from zero, is localized in the vicinity of the
detachment height $h_D$. The corresponding theoretical prediction for the
relatively weak adsorption strength is depicted in Fig. \ref{pdf_theory}b. It
can be seen that with increasing $h$ the almost Gaussian distribution tends to
Poisson-like one. Also the fluctuations decrease with $h$ in accordance with
MC-findings.

\begin{figure}[bht]
\hspace{-2.0cm}
\includegraphics[scale=0.30]{f_h_plateau.eps}
\hspace{0.50cm}
\includegraphics[scale=0.40]{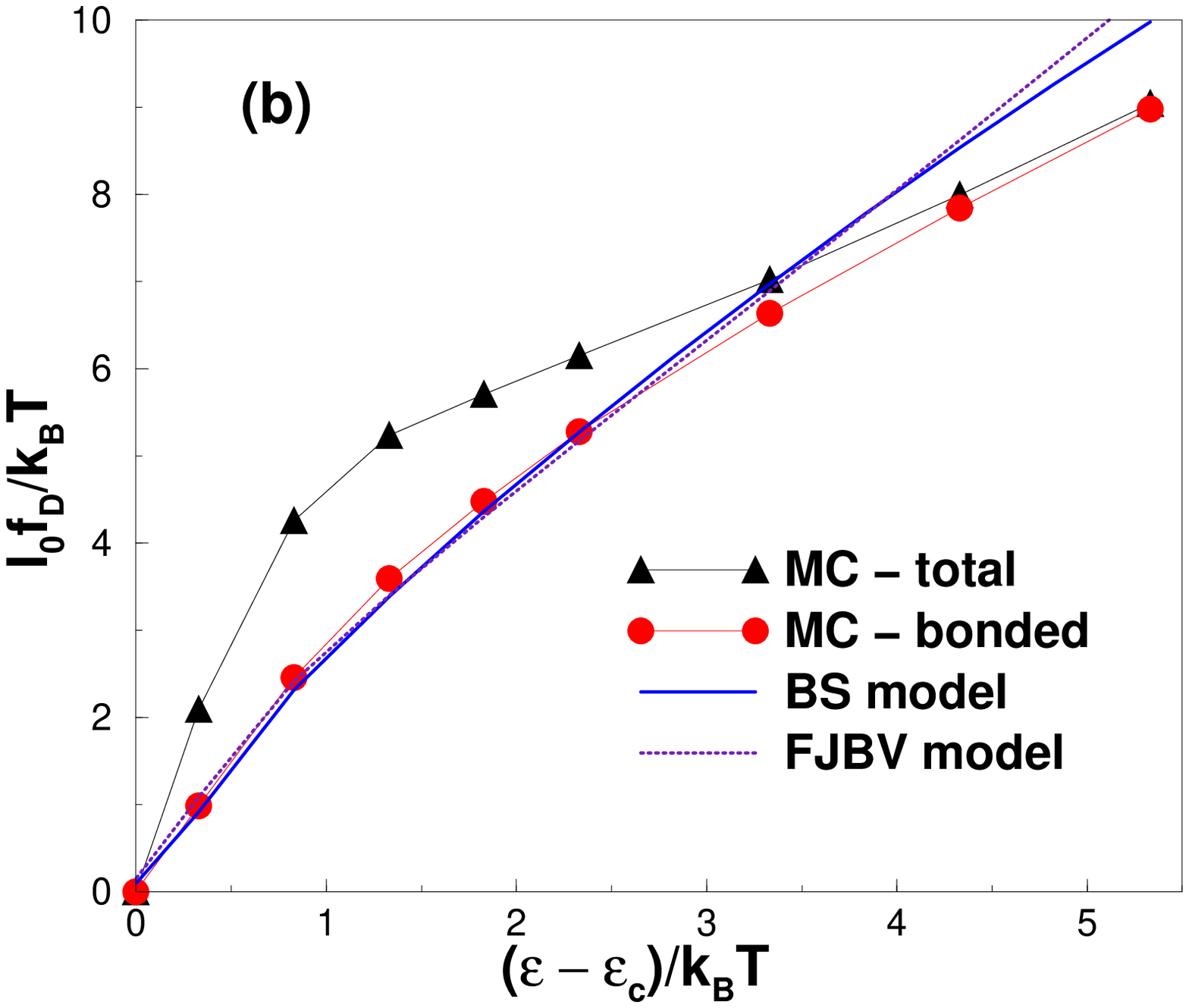}
\caption{\label{force} (a) Variation of the two components to the total force,
exerted by the chain on the end-monomer which is fixed at (dimensionless) height
$h/l_0N$ for different adsorption potentials $2.0 \le \epsilon/k_BT \le 5.0$:
bonding interactions (full symbols) and non-bonding Morse interactions (empty
symbols). In the inset the same is shown for a neutral plane $\epsilon = 0.0$
for purely repulsive monomers (triangles) and for such with a weak Morse
attraction (circles). (b) Variation of the total force (plateau hight) exerted
by the AFM tip on the chain-end for chain length $N=128$ with the
relative strength of adsorption $(\varepsilon - \varepsilon_c)/k_BT$.}
\end{figure}

The force $f$, exerted by the chain on the end-monomer, when the latter is kept
at height $h$ above the surface, is one of the main properties which can be
measured in experiments carried out within the fixed-height ensemble. Note that
$f$ has the same magnitude and opposite sign, regarding the force, applied by
the experimentalist. The variation of the force $f$ with increasing height $h$
is shown in Fig.~\ref{force}a for several values of the adsorption potential
$2.0 \le \epsilon \le 5.0$. In Fig.~\ref{force}a we distinguish between two
contributions to the total force $f$, acting on the end bead. The first stems
from the quasi-elastic forces of the bonded interaction (FENE) whereas the
second contribution is due to the short-range (attractive) interactions between
non-bonded monomers (in our model - the Morse potential). A typical feature of
the $f - h$ relationship, namely, the existence of a broad interval of heights
$h$ where the force remains constant (a plateau in the force) is readily seen in
Fig.~\ref{force}a. With growing strength of adsorption $\epsilon$ the length of
this plateau as well as the magnitude of the plateau force increase. Note, that
for $\epsilon = 0$ no plateau whatsoever is found. Upon further extension (by
increasing $h$) of the chain, the plateau ends and the measured force starts
to grow rapidly in magnitude - an effect, caused by a change of the chain
conformation itself in the entirely desorbed state.

A closer inspection of Fig.~\ref{force}a reveals that the non-bonded
contribution to $f$, which is generally much weaker than the bonded one,
behaves differently, depending on whether the forces between non-nearest
neighbors along the backbone of the chain are purely repulsive, or contain an
attractive branch. While for strong adsorption, $\epsilon \ge 3.0$, a plateau
is observed even for attractive non-bonded interactions, for weak adsorption,
$\epsilon \le 2.0$, an increase of the non-bonded contribution at $h/l_0N
\approx 0.35$, (seen as a {\em minimum} in Fig.~\ref{force}a) is observed. This
effect is entirely missing in the case of purely repulsive nonbonded
interactions - see the inset in Fig.~\ref{force}a where the contributions
from bonded and non-bonded interactions are shown for a neutral surface
$\epsilon = 0$. If one plots the magnitude of the measured force at the
plateau against the corresponding value of the the adsorption potential,
$\epsilon$, one may check the theoretical result, Eq.~(\ref{Local_Equilibrium})
- Fig.~\ref{force}b. Evidently, the theoretical predictions about $f_{\rm p}$ agree
well with the observed variation of the detachment force in the MC simulation,
both within the BS- or FJBV models, as long as only the excluded volume
interactions in the MC data are taken into account. If the total contribution to
$f_{\rm p}$, including also attractive non-bonded interactions in the chain, is
depicted - black triangles in Fig.~\ref{force}b - then the agreement with the
theoretical curves deteriorates since the latter do not take into account the
possible presence of attractive non-bonded interactions.
\vspace{0.6cm}
\begin{figure}[bht]
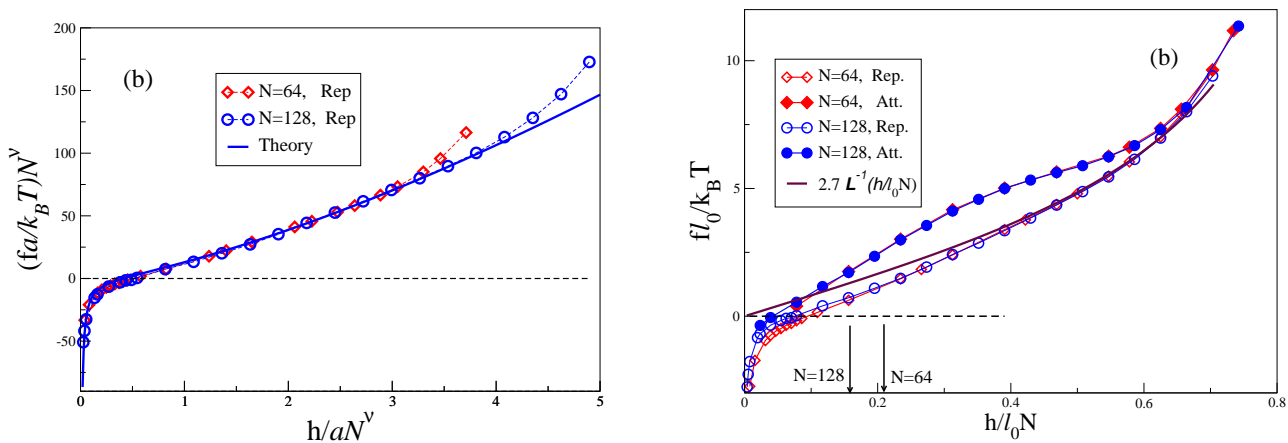

\hspace{-2.0cm}
\includegraphics[scale=0.32]{h_f_Pincus.eps}
\hspace{0.90cm}
\includegraphics[scale=0.31]{force_B.eps}
\caption{\label{f-h} (a) Variation of the total applied force $f$ with growing
height of the end monomer in terms of Pincus reduced variables, $f l_0
N^{\nu}/k_B T$ versus  $h/l_0N^\nu$,  for a polymer with purely repulsive
nonbonded forces for $N=64,\;128$. (b) The same as in (a)  but in terms of
reduced units $f l_0/k_B T$ versus  $h/l_0N$ for purely repulsive (empty
symbols)
as well as  for usual Morse potential  (full symbols) of nonbonded interactions
between monomers. The FJBV-model  results, Eq.~(\ref{L}), is
shown by a solid line. Arrows indicate the unperturbed gyration radius positions
$R_g/N$ for $N=64,\;128$}
\end{figure}

The $f - h$ relationship, which gives the equation of state of the stretched
polymer, may be derived within one of the different theoretical models, e.g.,
that of BS-, Eq.~(\ref{Deformation}), or FJBV-model, Eq.~(\ref{L}), as
mentioned in Section \ref{Theory}. Which of these theoretical descriptions is
the more adequate can be decided by comparison with experiment. In
Fig.~\ref{f-h}a,b, we present such comparison by plotting our simulation data
using different normalization for the height $h$. From Fig.~\ref{f-h}a it
becomes evident that the data from our computer experiment for $N=64$ and
$N=128$ collapse on a single curve, albeit this collapse only holds as long as
$h/l_0N^\nu \le 3.0$ for the BS-model while it fails for stronger
stretching. In contrast, this collapse works well for all values of $h$,
provided the height is scaled with the contour length of the chain $N$, rather
 than with $N^\nu$, as in the FJBV model - Fig.~\ref{f-h}b - regardless of
whether a purely repulsive, or the full Morse potential (which includes also an
attractive part) of interactions is involved. The analytical expression,
Eq.~(\ref{L}), is found to provide perfect agreement with the simulation data
for strong stretching, $h/l_0N \ge 0.4$. From the simulation data on
Fig.~\ref{f-h} one may even verify that the force $f$ goes through zero at some
height $h > 0$ and then turns negative, provided one keeps the chain end very
close to the grafting surface (cf. eq.(\ref{Deformation})).

\section{Summary}

In the present work we have treated the force-induced desorption of a
self-avoiding polymer chain from a flat structureless substrate both
theoretically and by means of Monte Carlo simulation within the constant-height
ensemble. The motivation for this investigation has been the necessity to
distinguish between results obtained in this ensemble and results, derived in
the constant-force ensemble, considered recently\cite{Bhattacharya}, as far as
both ensembles could in principle be used by experimentalists. We demonstrate
that the observed behavior of the main quantity of interest, namely, the
fraction of adsorbed beads $n$ (i.e., the order parameter of the phase
transition) with changing height $h$ differs qualitatively from the variation of
the order parameter when the pulling force is varied. In the constant-height
ensemble one observes a steady variation of $n$ with changing $h$ whereas in
the constant-force ensemble one sees an abrupt jump of $n$ at a particular
value of $f_D$, termed a detachment force. However, this should not cast doubts
on the genuine first-order nature of the phase transition which can be recovered
within the constant-height ensemble too, provided one expresses the control
parameter $h$ in terms of the average force $f$. This equivalence has been
studied extensively for Gaussian chains by Skvortsov et al. \cite{Skvortsov} who
noted that ensemble equivalence does not apply to fluctuations of the pertinent
quantities too.

Indeed, in our earlier study\cite{PRER} we found diverging variance of the PDF
$P(n)$ at $f_D$ whereas in our present study the fluctuations of the order
parameter are observed to stay finite at the transition height $h_D$. These
findings confirm theoretical predictions based on analytic results which we
derive within the GCE-approach.  Within this approach we have explored two
different theoretical models for the basic force - extension relationship,
namely, the bead-spring (BS) model as well as that of a Freely-Jointed
Bond-Vectors (FJBV) model. Our simulation results indicate a good agreement
between theory and computer experiment.

\section*{Acknowledgments}
We are indebted to A. Skvortsov  for useful discussions during the preparation
of this work. A.~Milchev thanks the Max-Planck Institute for Polymer Research in
Mainz, Germany, for hospitality during his visit at the institute. A.~Milchev
and V.~Rostiashvili acknowledge support from the Deutsche Forschungsgemeinschaft
(DFG), grant No. SFB 625/B4.

\begin{appendix}
 \section{Freely jointed bond vectors model}
The deformation law in the  overstretched regime (when the chain deformation
is close to its saturation) could be treated better within the FJBV model.
Consider a tethered chain of length $N$ with one end anchored at the
origin of the coordinates and an external force $f_N$ acting on the free end of
the chain. The corresponding deformation energy reads
\begin{eqnarray}
 U_{\rm ext } = - f_N \: r_{N}^{\perp} = - f_N \: \sum_{i=1}^{N} \: b_i
\cos \theta_i
\end{eqnarray}
where $r_{N}^{\perp}$ is the $z$-coordinate (directed perpendicular to the surf
ace) of the chain end, $b_i$ and $\theta_i$ are  the length  and the polar angle
of the $i$-th bond vector respectively. The corresponding partition function of
the FJBV model is given by
\begin{eqnarray}
 Z_{N} (f_N) &=& \int \prod_{i=1}^{N} d \phi_i \sin \theta_i d \theta_i
\exp\left(\frac{f_N}{k_B T} \sum_{i=1}^{N} b_i \cos \theta_i\right)\nonumber\\
&=& (4\pi)^{N} \prod_{i=1}^{N}  \left(\frac{k_B T}{b_i f_N} \right) \cosh
\left(\frac{b_i f_N}{k_B T} \right)
\label{Partition_Func}
\end{eqnarray}
The average orientation of the $i$-th bond vector can be calculated as
\begin{eqnarray}
 <\cos \theta_i> = \left(\frac{k_B T}{f_N}\right) \:   \frac{\partial}{\partial
b_i}
\ln Z_N (f_N) = \coth \left(\frac{b_i f_N}{k_B T}\right) - \left(\frac{k_B
T}{b_i f_N
}\right)
\label{Cos}
\end{eqnarray}
From Eq.(\ref{Cos}) the chain end mean distance from the surface, $h$, is given
by
\begin{eqnarray}
h = \sum_{i=1}^{N} b_i <\cos \theta_i> = b N {\cal L} \left(\frac{b f_N }{k_B T}
\right)
\label{H_vs_Force}
\end{eqnarray}
where we have taken into account that the lengths of all bond vectors are of
equal length, $b_i = b$, and ${\cal L} (x) = \coth (x) - 1/x$ is the Langevin
function. This leads to the force - distance relationship
\begin{eqnarray}
 \frac{b f_N}{k_B T} = {\cal L}^{-1}\left(\frac{h}{b N}\right) = \begin{cases}
                                                              \frac{1}{1-h/b
N} & \text{, at}  \:\: h/b N \leq 1\\
\frac{2 h}{b N}  & \text{, at} \:\:  h/b N \ll 1
                                                             \end{cases}
\label{Inversion}
\end{eqnarray}
which we use in Sec.II. The notation ${\cal L}^{-1}(x)$ stands for the inverse
Langevin function.

\section{Calculation of PDF in the $h$-ensemble}

Using Eq.(\ref{Z_Star}) for $\ln z^{*} (w)$ as well as Eqs.
(\ref{Local_Equilibrium}) and (\ref{M_vs_h}) (for the BS-model), one obtains
an expression for the tail length
\begin{equation}
 M(h, \xi \: w) = \dfrac{h/l_0}{\left[a_1^\phi (w - w_c)
\: \xi \right]^{(1-\nu)/\phi} }
\label{Tail_Length}
\end{equation}
The saddle point (SP) equation in this case reads (cf. Eq.(\ref{SP}))
\begin{equation}
 \dfrac{K+1}{\xi_0} = - \left. \left[N - M(h, \xi_0 w) \right]   \:
\dfrac{\partial \ln z^{*}}{\partial
\xi}\right|_{\xi = \xi_0} + \left. \dfrac{\partial M}{\partial \xi} \right|_{\xi
= \xi_0} \: \ln z^{*}(\xi_0 w)
\label{SP_General}
\end{equation}
Taking into account Eqs. (\ref{Z_Star}) and (\ref{Tail_Length}), after
introducing the notation $y = (w - w_c) \xi_0$, the SP-equation can be recast
into
 \begin{equation}
\left( \dfrac{K}{N}\right) = y^{1/\phi}  - B_1 \left( \dfrac{h}{l_0 N}\right)
y^{\nu/\phi} - B_2 \left( \dfrac{h}{l_0 N}\right) \dfrac{\ln
\mu_3}{y^{(1-\nu)/\phi}}\;,
\label{SP_Recast}
 \end{equation}
where $B_1$ and $B_2$ are constants of the order of unity. In the particular
case $h = 0$ Eq.(\ref{SP_Recast})  goes back, as expected, to
Eq.(\ref{SP_Solution}). Eq. (\ref{SP_Recast}) can be solved iteratively for $h
\ll l_0 N$ as
 \begin{equation}
  y = \left[ \left( \dfrac{K}{N}\right)  + B_1 \left( \dfrac{K}{N}\right)^\nu
\left( \dfrac{h}{l_0 N}\right) + B_2 \ln \mu_3 \left(
\dfrac{N}{K}\right)^{1-\nu} \left( \dfrac{h}{l_0 N}\right) \right]^{\phi}
\label{Iteration}
 \end{equation}
As before, the main contribution in the integral given by  Eq.
(\ref{Coefficient}) reads
 \begin{eqnarray}
P(K) &\propto& \exp \left\lbrace- \left[N - M(h, \xi_0 w)\right] \ln
z^{*}(\xi_0 w) - (K+1) \ln \xi_0 \right\rbrace \nonumber\\
&=& \left\lbrace (w -
w_c) N^{\phi} \exp \left[1 - \dfrac{(h/l_0 N)}{(K/N)^{1-\nu} \left[\rho
(K/N, h/l_0 N)\right]^{1-\nu}}\right] \rho (K/N, h/l_0 N)\right\rbrace^K \:
\left[K \rho (K/N, h/l_0 N) \right]^{-\phi K}
 \end{eqnarray}
where one introduces the notation
\begin{eqnarray}
 \rho \left(K/N, h/l_0 N \right) \equiv 1 + B_1 \dfrac{h/l_0
N}{(K/N)^{1-\nu}} + B_2 \ln \mu_3 \: \dfrac{(h/l_0 N)}{(K/N)^{2-\nu}}
\label{Notation}
\end{eqnarray}
After normalization, the final expression for the PDF reads
\begin{eqnarray}
 P(K) = \dfrac{1}{{\cal W}(\eta, h/l_0 N)} \: \left\lbrace \eta \: \exp
\left[1 - \dfrac{(h/l_0 N)}{(K/N)^{1-\nu} \left[\rho
(K/N, h/l_0 N)\right]^{1-\nu}}\right] \rho (K/N, h/l_0 N)\right\rbrace^K \:
\left[K \rho (K/N, h/l_0 N) \right]^{-\phi K}
\label{PDF_Pull}
\end{eqnarray}
where $\eta = (w - w_c) N^\phi$  as before and the normalization constant reads
\begin{eqnarray}
 {\cal W}(\eta, h/l_0 N) = \sum_{K=1}^{N} \: \left\lbrace \eta \: \exp
\left[1 - \dfrac{(h/l_0 N)}{(K/N)^{1-\nu} \left[\rho
(K/N, h/l_0 N)\right]^{1-\nu}}\right] \rho (K/N, h/l_0 N)\right\rbrace^K \:
\left[K \rho (K/N, h/l_0 N) \right]^{-\phi K}
\label{Normalization_Pull}
\end{eqnarray}
Again one can readily see that for $h = 0$ Eq.(\ref{PDF_Pull}) reduces to
Eq.(\ref{Distr3}). The PDF, following from Eq.~(\ref{PDF_Pull}) is shown in
Fig.~\ref{pdf_theory}b for several values of the height $h$. Evidently, both
the mean value $\overline{K}$ and the variance $\overline{K^2}-\overline{K}^2$
decline with growing $h$.

\end{appendix}

\end{document}